\documentclass[openany]{nature}
\usepackage[T1]{fontenc}
\usepackage{amsthm,amsmath,amssymb}
\usepackage{mathrsfs}
\usepackage{array}
\usepackage{overpic}
\usepackage[T1]{fontenc}
\usepackage{graphicx}
\usepackage{pdflscape}
\usepackage{hyperref}
\hypersetup{hidelinks} 
\usepackage{threeparttable}
\usepackage{xcolor}
\usepackage{ulem}
\usepackage{cancel}
\usepackage[figuresleft]{rotating}
\usepackage{lineno}
\usepackage{caption}
\usepackage{subfigure}
\usepackage{subcaption}
\usepackage{tablefootnote}
\usepackage{amsmath, amssymb}

\makeatletter

\let\saved@includegraphics\includegraphics
\AtBeginDocument{\let\includegraphics\saved@includegraphics}

\makeatother
\usepackage{aas_macros}

\def\be{\begin{eqnarray}}
\def\ee{\end{eqnarray}}

\makeatletter

\usepackage{graphicx}
\usepackage{longtable}
\usepackage{supertabular}
\usepackage{float} 
\usepackage{authblk}
\usepackage{CJK}
\usepackage{orcidlink}

\title{Probable evidence for a transient mega-electron volt emission line in the  GRB~221023A}

\author{Lu-Yao Jiang$^{1,2}$\orcidlink{0000-0002-2277-9735}, Yun Wang$^{1}$\orcidlink{0000-0002-8385-7848},  
Yu-Jia Wei$^{1,2,3,4}$\orcidlink{0000-0002-9775-2692}, 
Da-Ming Wei$^{1,2}$\thanks{E-mail:dmwei@pmo.ac.cn}   \orcidlink{0000-0002-9758-5476}, 
Xiang Li$^{1,2}$\orcidlink{0000-0002-5894-3429},  
Hao-Ning He$^{1,2}$\orcidlink{0000-0002-8941-9603}, 
Jia Ren$^{1}$\orcidlink{0000-0002-9037-8642}, 
Zhao-Qiang Shen$^{1}$\orcidlink{0000-0003-3722-0966}, 
Zhi-Ping Jin$^{1,2}$\orcidlink{0000-0003-4977-9724}.}

\begin{document}
\maketitle
\begin{affiliations}
\item{Key Laboratory of Dark Matter and Space Astronomy, Purple Mountain Observatory, Chinese Academy of Sciences, Nanjing 210023, China}
\item{School of Astronomy and Space Science, University of Science and Technology of China, Hefei 230026, China}
\item{Department of Astronomy and Astrophysics, 525 Davey Lab, The Pennsylvania State University, University Park, PA 16802, USA}
\item{Institute for Gravitation and the Cosmos, The Pennsylvania State University, University Park, PA 16802, USA}
\end{affiliations}

\section*{Abstract}
\begin{abstract} 
Detection of spectral line in gamma-ray bursts (GRBs) is importance for studying GRB physics, as it provides insights into the composition and physical conditions of the GRB environment. However, progress in detecting X-ray or gamma-ray emission and absorption lines in GRB spectra has been relatively slow, only the narrow emission line feature of about 10 MeV found in GRB~221009A has exhibited a significance exceeding $ 5 \sigma$. Here, we report the probable evidence of a narrow emission feature at about 2.1 mega–electron volts (MeV) in the spectrum of GRB 221023A. The highest statistical significance of this feature is observed in the time interval between 8 and 30 seconds after Fermi Gamma-Ray Burst Monitor trigger, with the chance probability value $<2.56\times10^{-5}$ (after accounting for the look-elsewhere effect), corresponding to a Gaussian-equivalent significance$> 4.20\sigma$. We interpret this feature as being generated through the de-excitation of excited electrons in the relativistic hydrogen-like high-atomic-number ions entrained in the GRB jet. 
\end{abstract}

\section*{Introduction}
Gamma-ray bursts (GRBs) are the most luminous stellar explosions in the universe. These events generally appear as brief and intense $\gamma$-rays followed by a long-lived afterglow emission. The GRB prompt emission originates from relativistic jets that dissipate the energy and accelerate particles either via internal shocks or magnetic reconnection, with high variability and usually lasts from milliseconds to thousands of seconds\cite{Rees1994, ZhangBing2011, Peer2015}. Most of the observed spectrum of GRB prompt emission in the keV to MeV energy range usually can be described by a smoothly joint broken power-law function (called the Band function\cite{Band1993}). Despite decades of intensive investigation, our understanding of the physics behind the prompt emission of GRBs remains limited.

The existence of X-ray or gamma-ray emission and absorption lines in the GRB energy spectrum has been debated. For example, in prompt emission phase, the Konus instrument detected absorption lines at  $\rm 30-70 \ keV$  and emission lines at $\rm 400-460 \ keV$ in the energy spectra of some GRBs\cite{Mazets1980, Mazets1981}. HEAO-1 observed  absorption-like features in the spectra of some GRBs\cite{Hueter1982, Hueter1984}. The Japanese Ginga Gamma-Ray Burst Detector (GBD) observed two absorption-like features in three GRBs (GRB~870303, GRB~880205 and GRB~890929), which may be interpreted as the first and second cyclotron absorption lines\cite{Murakami1988, Yoshida1991, Yoshida1992}.  Additionally, some $6.4 \ \rm keV$ iron K-$\alpha$ spectral lines were claimed to have been found in some bursts. Within the energy range of $3.8\pm0.3\rm \ keV$, a possible transient Fe absorption feature was identified in the prompt X-ray spectrum of GRB~990705. This feature appeared during the initial rising phase of the burst profile and disappeared thereafter\cite{Amati2000}. In another study, Frontera et al.\cite{Frontera2004} analyzed the prompt emission spectrum of GRB~011211 and found potential indications of transient Fe absorption features around $\rm 6.9\pm0.6 \ keV$ during the rise of the main pulse.  
However, the statistical significance of these features  is found to be below the $5\sigma$ threshold. Even when extending the spectral lines search to the afterglow phase and  conducting large-scale searches using X-ray detection satellites such as Chandra\cite{2000SPIE.4012....2W}, Swift X-ray Telescope\cite{2005SSRv..120..165B}, X-ray Multi-Mirror Mission Newton (XMM-Newton)\cite{2001A&A...365L...1J},  Advanced Satellite for Cosmology and Astrophysics (ASCA)\cite{1994PASJ...46L..37T}, Satellite per Astronomia X (BeppoSAX)\cite{1997A&AS..122..299B},  no credible X-ray line feature has been detected in GRBs afterglow\cite{Sako2005, Hurkett2008, Campana2016}.

More recently, a highly significant ($>5\sigma$)  narrow emission feature around 10~MeV has been detected in the Fermi data of GRB~221009A \cite{Edvige2023, Zhang2024obs}. These intriguing features appear  during the decay phase of the brightest pulse, with the central energy of the Gaussian distribution gradually  shifting   towards lower energies over time (about $37 \ \rm MeV$ to $6 \ \rm MeV$), while the ratio of the line width to the central energy is nearly constant (about $10 \%$). At the same time, in the Konus-Wind data of GRB~221009A, a similar narrow emission feature has been found  with a significance level below $2\sigma$~\cite{Frederiks2023}. Two independent satellites simultaneously detected the narrow emission feature in GRB 221009A, further bolstering the credibility of the narrow emission feature observed in GRB~221009A. 

In this work, we perform a spectral analysis of the prompt emission from GRB~221023A using Fermi Gamma-Ray Burst Monitor (GBM)\cite{Meegan2009} data. We find a marginally significant narrow emission feature around 2.1~MeV. The highest statistical significance of this feature is observed in the time interval $8-30 \ \rm s$, with the chance probability value $<2.56\times10^{-5}$ (after accounting for the look-elsewhere effect), corresponding to a Gaussian-equivalent significance$> 4.20\sigma$. We find that the relativistic hydrogen-like high-atomic-number ions entrained in the GRB jet can generate such narrow MeV emission lines through the de-excitation of excited electrons.

\section*{Results}
\subsection{Light curve and spectral analysis.}
GRB~221023A triggered the GBM onboard Fermi at 20:41:34.92 UT on 23 October 2022~\cite{Dunwoody2022}. Simultaneously,  this event was also detected by Konus-Wind \cite{Ridnaia2022} and AGILE (Astrorivelatore Gamma ad Immagini LEggero) \cite{Ursi2022}.  The GBM light curve shows one bright peak with a total duration time $T_{90}$ about 39 seconds (s) in the $50-300 \ \rm keV$ energy band \cite{Dunwoody2022}, and the fluence reported in the Fermi-GBM catalog is $F=3.41 \times 10^{-4} \rm  \ erg \ cm^{-2}$ in the energy range $\rm 10-1000 \ keV$ \cite{vonKienlin2020ApJ}. The Fermi Large Area Telescope (LAT)\cite{2009ApJ...697.1071A} instrument was triggered during this event, and the highest-energy photon detected is a 17 GeV event with a $99\%$ probability which is observed 576 seconds after the GBM trigger \cite{Pillera2022}.

\begin{figure*}
\centering
\includegraphics[width=1.0\hsize]{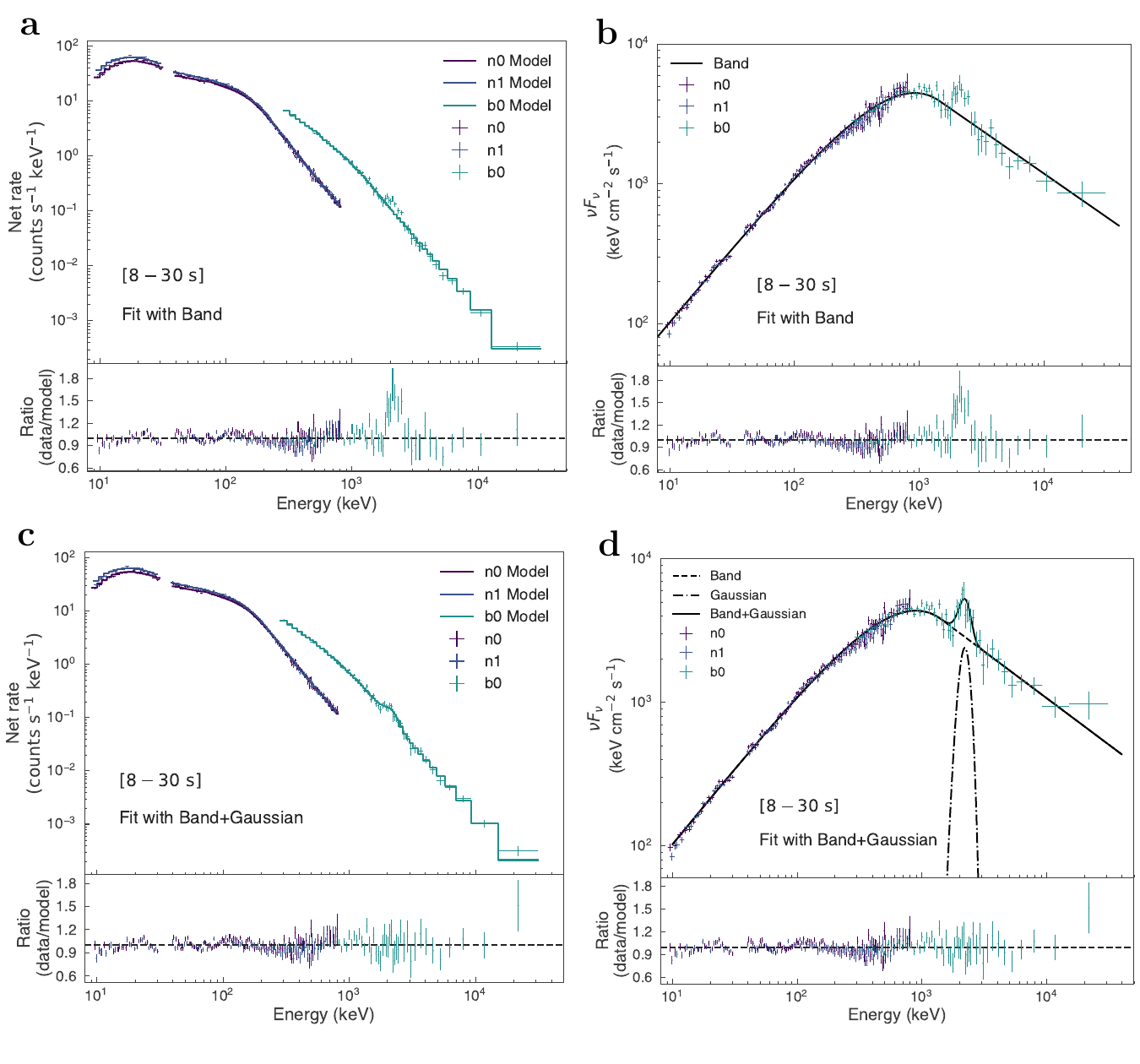}
\caption{\textbf{Spectral fitting within the $8-30 \ \rm s$ time interval.}  The counts rate spectrum in the panel \textbf{a} and  the $\nu F_{\nu}$ spectrum in the panel \textbf{b} are obtained from fitting the Band function. Data are from GBM’s two sodium iodide (NaI) detectors (n0: dark purple, n1: blue) and one BGO detector (b0: light cyan-green).
The narrow feature appears as an excess around  $1 \rm \ MeV-3\ \rm MeV$  in the b0 detector data.  Panels \textbf{c} and \textbf{d} show the same spectra fitted with the Band function plus a Gaussian component to model the observed excess. Error bars indicate the $1\sigma$ uncertainty on data points. Source data are provided as a Source Data file.}
\label{fig:1}
\end{figure*}

\begin{figure*}
\centering
\includegraphics[width=0.8\hsize]{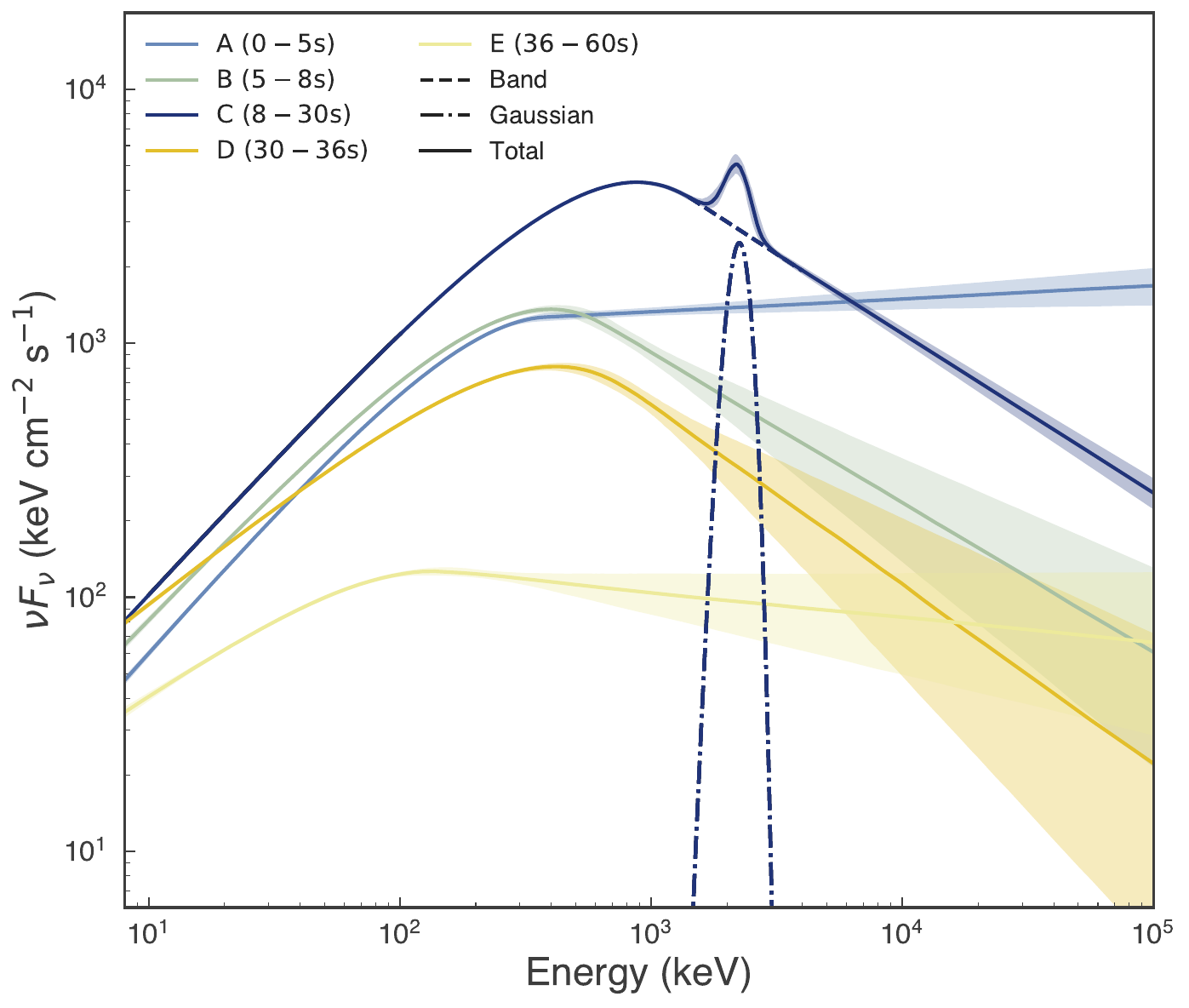}
\caption{\textbf{Energy spectrum evolution.} Best-fit $\nu F_{\nu}$ model spectra for the time-resolved data in different time intervals, five time intervals are color-coded, with the corresponding shaded colors show the $68\%$ confidence levels.}
\label{fig:evolution}
\end{figure*}

Panels a, b of Figure~\ref{fig:lightcure} presents light curves for GRB~221023A at different energy bands. We analyzed the spectral evolution of the GRB prompt emission by spectra in 5 adjacent time intervals (labelled A~($0-5 \ \rm s$), B ($5-8 \ \rm s$), C ($8-30 \ \rm s$), D ($30-36 \ \rm s$) and E ($36-60 \ \rm s$)). The spectra of  time intervals $0-5 \ \rm s$, $5-8 \ \rm s$, $30-36 \ \rm s$ and $36-60 \ \rm s$ can be fitted with a Band function (see methods subsection Spectral fitting), detailed analysis results are summarized in Table~\ref{tab:1}. Interestingly, when fitting the spectra in time interval $8-30 \ \rm s$ using the Band function,  as shown in the a and b panels of Figure~\ref{fig:1}, revealing a distinct narrow and bright emission feature between 1~MeV and 3~MeV. This narrow emission feature can be well modeled by adding a Gaussian component on top of the Band function (see methods subsection Spectral fitting), the best-fit parameter values are $\alpha=-0.93_{-0.01}^{+0.01}$, $E_{p}=891.07_{-33.19}^{+3.03} \ \rm keV$, $\beta=-2.65_{-0.02}^{+0.05}$, $E_{\rm gauss}=2154.60_{-65.07}^{+53.37} \ \rm keV$, $\sigma_{\rm gauss}=229.36_{-45.29}^{+93.57} \ \rm keV$, the c and d panels of Figure~\ref{fig:1} displays the corresponding fitted counts rate and $\nu F_{\nu}$ spectrum. We also present the best-fit $\nu F_{\nu}$ model spectra in 5 adjacent time intervals in Figure~\ref{fig:evolution}.
In time interval $8-30 \ \rm s$, comparing models with and without the Gaussian component, we obtained $\Delta \rm AIC=51.87$, $\rm ln(BF)=9.99$, and $\Delta \chi^2 = 40.14$, which strongly supports the presence of an additional narrow emission feature (see methods subsection Model comparison). 
In order to exclude the impact of background subtraction on the extracted spectrum, we calculated the background  spectrum by selecting several different time windows. The result of the narrow emission feature is  substantially unaffected (see methods subsection Background).

We performed a time-resolved spectral analysis on time interval $8-30 \ \rm s$ to further investigate the presence of the observed feature and to characterize its evolution. We used a fixed window size of $13 \ \rm s$, sliding it in steps of $3 \ \rm s$ to divide the time intervals, resulting in four subintervals, referred to as $8-21 \ \rm s$ (C.1), $11-24 \ \rm s$ (C.2), $14-27 \ \rm s$ (C.3), $17-30 \ \rm s$ (C.4). In four subintervals, we still extracted the spectra by performing a different selection of the time windows for the background spectrum computation (see methods subsection Background). The narrow emission feature remains clearly visible in these finer time intervals. 
In the four finer time-resolved spectra  ($8-21 \ \rm s$, $11-24 \ \rm s$, $14-27 \ \rm s$, $17-30 \ \rm s$), the $\Delta \rm AIC$ values vary between $25.76-36.55$, the $\rm ln(BF)$ ranges from  $2.06$ to $7.34$, and the $\Delta \chi^2$ ranges from $18.53$ to $34.49$. These results further strongly favor  adding an additional narrow emission feature (see methods subsection Model comparison).
The comprehensive results of the spectral analysis for these four time intervals are presented in Table~\ref{tab:1}. The temporal evolution of Gaussian component parameters is presented  in the (f), (g), and (h) panels of Figure~\ref{fig:lightcure}. Notably,   the central energy $E_{\rm gauss}$ of the narrow emission feature remains constant at around $2.1 \ \rm MeV$, while the width $\sigma_{\rm gauss}$ shows a possible decreasing trend over time. The flux of the narrow emission feature is approximately $10^{-6} \ \rm erg \ cm^{-2} \ s^{-1}$.

\subsection{Significance analysis.}
In order to assess the significance of narrow emission features, we created $1.00\times10^{7}$ simulated data assuming the Band model and fitted them with both the Band and Band+Gaussian models to obtain the distribution of $\Delta \chi^2$. Table \ref{tab:2} shows the chance probability values ($p \text{-} {\rm value}_{\rm sim}$) calculated based on the results of $1.00\times 10^7$ simulations. We also considered the $p \text{-} {\rm value}_{\rm sim}$ corrected for the number of independent search trials ($p \text{-} {\rm value}_{\text{sim-trial}}$).  The highest statistical significance of narrow emission features was observed in the time interval C ($8-30 \ \rm s$), with the chance probability value $p \text{-} {\rm value}_{\rm sim}<1\times10^{-7}$ obtained from results of $1\times 10^7$ simulations, corresponding to a Gaussian-equivalent significance $> 5.32\sigma$. Considering the correction for the number of independent search trials, the chance probability value decreases to $p \text{-} {\rm value}_{\text{sim-trial}}<2.56\times10^{-5}$, corresponding to a Gaussian-equivalent significance $> 4.20\sigma$ (see methods subsection Significance calculation of narrow emission feature). 
The chance probability values for the other time intervals are shown in Table \ref{tab:2}.

\subsection{Comparison with GRB~221009A.}
\begin{figure*}
\centering
\includegraphics[width=1\hsize]{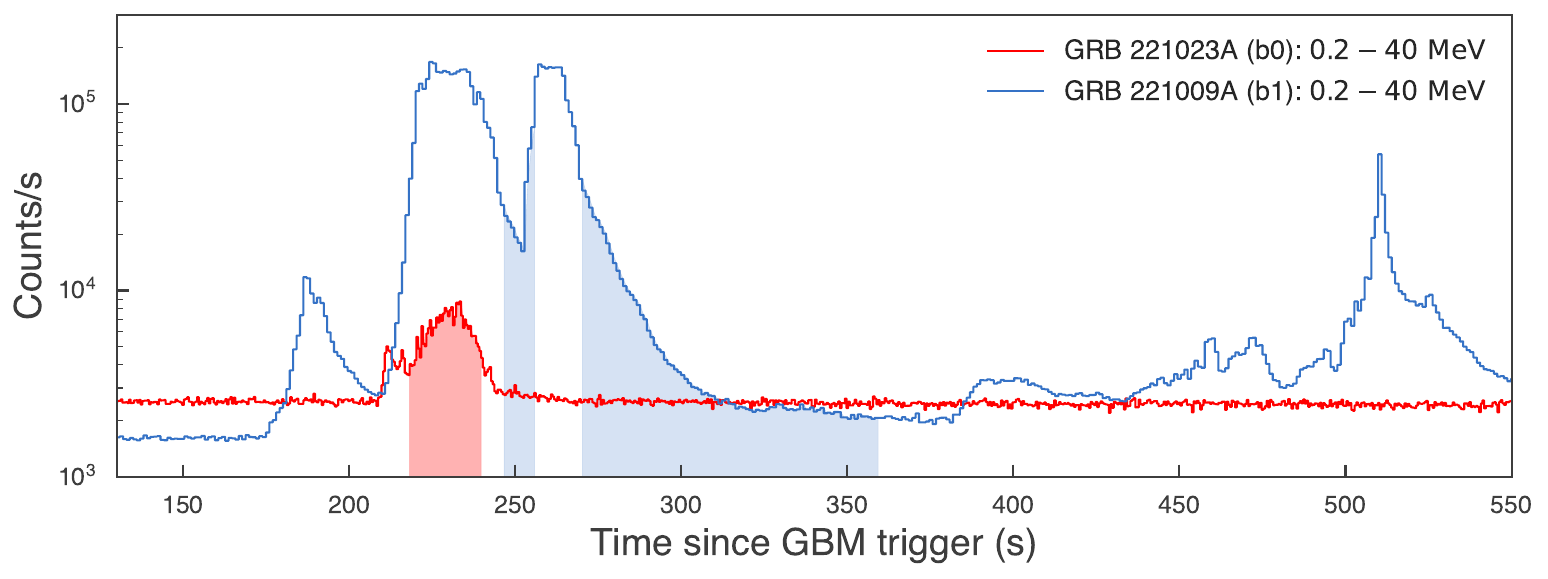}
\caption{\textbf{Comparison of light curve.} Comparison of the light curves of GRB~221023A (red) and GRB~221009A (blue) in the energy range of $\rm 0.2-40 \ MeV$, with the corresponding shaded colors regions showing the time intervals where narrow emission features were detected. The trigger time of GRB~221023A was shifted backward by $210 \rm \ s$. Source data are provided as a Source Data file.
}
\label{fig:comparison}
\end{figure*}

The spectral analysis of GRB~221023A reveals a  marginally statistically significant narrow emission feature at around 2.1~MeV. This would then represent the second event following GRB 221009A with a narrow emission feature in the MeV energy range. In the case of GRB~221009A,  the central energy $E_{\rm gauss}$ of the narrow emission  feature decreases over time (about $37 \ \rm MeV$ to $6 \ \rm MeV$),  while the ratio of the line width to the central energy is nearly constant (about $10\%$)~\cite{Edvige2023, Zhang2024obs}. For GRB~221023A, we observe the trend: the central energy $E_{\rm gauss}$ remains steady at around $2.1 \ \rm MeV$ throughout the observation period, while the width $\sigma_{\rm gauss}$ exhibits a possible decreasing trend as time progresses, the flux of the narrow
emission feature is around $10^{-6} \ \mathrm{erg \ cm^{-2} \ s^{-1}}$.  
The  Figure~\ref{fig:comparison} displays the lightcurves of GRB~221009A and GRB~221023A within the energy range of $\rm 0.2-40 \ MeV$. The shaded regions indicate time intervals  in which  narrow emission  features were detected. The narrow emission feature in GRB~221023A appears during the rising and falling phases of the brightest pulse, with a duration of  $22 \ \rm s$ (time intervals: $8-30 \rm \ s$) and then disappears. In contrast,  the narrow emsission  feature in GRB~221009A appears during the falling phase of the brightest pulse, with a duration of  $100 \ \rm s$ (time intervals: $246-256 \rm \ s$ and $270-360 \rm \ s$) \cite{Edvige2023,Zhang2024obs}. Interestingly, the Fe absorption feature previously identified during the prompt emission of GRB~990705 and GRB~011211 appears during the rising phase of the main pulse \cite{Amati2000, Frontera2004}. This implies a higher likelihood of detecting emission or absorption features during  time intervals near the peak of the main pulse in the prompt emission phase of GRBs. It is worth noting that due to the very high photon flux of GRB~221009A, the Fermi-GBM experienced Bad Time Interval (time interval affected by saturation) between 219 and 277 seconds~\cite{Lesage2023}. It is possible that narrow emission features also exist during the rising phase of the main emission in GRB~221009A.

\section*{Discussion}
In general, standard models of prompt emission in GRBs do not predict the appearance of a transient  MeV narrow emission component\cite{rees94, Drenkhahn2002, ZhangBing2011}. To explain our potential finding, we have explored several possible scenarios. One possible explanation for the narrow emission feature is the blue-shifted annihilation line of relatively cold ($k_{B}T\ll m_{e}c^2$, where $k_B$ is Boltzmann constant, $T$ is the temperature of the medium, $m_e$ is the mass of the electron, and $c$ is the speed of light) electron-positron pairs. Within the emission region (resulting from internal shocks and/or magnetic reconnection) of GRB, electron-positron pairs are readily formed within the GRB jet (such as two-photon pair production $\gamma \gamma \longrightarrow e^{+} e^{-}$\cite{rees94, ZhangBing2011}). Numerical simulations of GRB spectra indicate that the generated spectra depend on the compactness of the fireball. In scenarios with high compactness, electron-positron pairs play an essential role in shaping the GRB prompt emission spectrum. In a pair-dominated fireball, a pair annihilation line is predicted\cite{Peer2004,Peer2006}. In the observer frame,  a line is expected to appear at an energy of $E_{\rm \pm,line}=\Gamma m_{e} c^2/(1+z)$,  where $\Gamma$ is the bulk Lorentz factor of the emitting region and $z$  represents the redshift. For the typical energy $E_{\rm line}$ is about $2.1 \rm \ MeV$ of the observed lines in GRB~221023A, the bulk Lorentz factor of the emitting region is required to be $\Gamma$ is about $4(1+z)$. Considering a redshift of $z=2$, the bulk Lorentz factor $\Gamma$ is about $12$. In this scenario, how such a low bulk  Lorentz factor is generated and maintained for an extended period in the prompt emission of the GRB is an issue.

The second scenario involves the possibility that the narrow emission feature is an intrinsic low-energy spectral line (such as the $6.4 \ \rm keV$ fluorescent  K-$\alpha$ iron line). This spectral line may be emitted within the region associated with the supernova ejecta. Subsequently, the energy of the spectral line could be boosted through up-scattering by the relativistic jet. The spectral line feature identified in GRB~221023A is narrow, which implies that electrons scattering photons are cold. This form of bulk Comptonization has already been proposed to occur within blazar jets\cite{Sikora1994}. The boosted photon energy of the low-energy spectral line is $E_{\rm line} = \Gamma^2E_{\rm low}/(1+z)$, where $\Gamma$ is the jet bulk Lorentz factor and $E_{\rm low}$ represents the low-energy spectral line of the particular element. If the observed spectral feature inGRB~221023A arises from the $6.4 \ \rm keV$ iron K-$\alpha$ line, and the typical photon energy of the observed line is around $2.1 \ \rm MeV$, this would require a jet bulk Lorentz factor $\Gamma$ to be about $18(1+z)^{\frac{1}{2}}$. Considering a redshift of $z=2$, the bulk Lorentz factor $\Gamma$ is about $31$. This scenario faces the same issue as the first one, namely, the low bulk Lorentz factor problem.

The third possible scenario is that the narrow emission feature may originate from  MeV nuclear de-excitation lines. The energetic particles interacting with ambient matter could excite heavy nuclei which can emit MeV  $\gamma$-ray line emissions via de-excitation, such as the 4.44 MeV line from $\rm ^{12}C$ and the 6.13 MeV line from $\rm ^{16}O$ \cite{Ramaty1979A, Murphy1987,Murphy2009}. In fact, nuclear de-excitation line emissions from $\rm ^{12}C$ and $\rm ^{16}O$ have been observed in solar flares\cite{Chupp1973, Chupp1975, Chambon1978}. Moreover, the existence of such nuclear de-excitation line has long been anticipated to be found within supernova remnants\cite{Summa2011, Weinberger2020, LiuBing2021, LiuBing2023}. The observed photon energy of the nuclear de-excitation line is  $E_{\rm line} = E_{\rm element} /(1+z)$, where $z$  is the redshift, and $E_{\rm element} $ corresponds to the energy of the particular element's nuclear de-excitation line.  If we assume that the narrow emission feature in GRB~221023A arises from nuclear de-excitation lines of $\rm ^{12}C$ or $\rm ^{16}O$, it corresponds to redshifts of $z = 1.1$ or $z = 1.9$, respectively.  
However, the radiation from the nuclear de-excitation of the ambient matter is almost isotropic, while the gamma-ray burst is collimated. Therefore, the energy budget needed to generate such a MeV emission line would be larger than the energy of the prompt emission for a typical half-opening angle $\theta_j = 0.1$~rad. Note that for a large $\theta_j$, such as $\theta_j > 0.3$~rad, this energy problem can be alleviated. 

The fourth possible scenario is that the heavy nuclei, especially relativistic hydrogen-like high-atomic-number ions originating from the $\beta$ decay of unstable nuclei and/or the recombination, entrained in GRB jets can produce such a narrow MeV emission line via electron de-excitation~\cite{Wei2024grb}. 
In this model, the reflection of the radiation from the WR star can generate enough seed photons to excite electrons.
After the jet with heavy nuclei breaks through the photosphere and hydrogen-like heavy ions are generated by $\beta$ decay and/or recombination, the emission line can be generated. 
The emission lines occur at an energy of $E_{\rm line} = \Gamma \epsilon_z / (1 + z)$, where $\epsilon_z = m_e \alpha^2 c^2 Z^2 / 2$ is the Rydberg energy, $\alpha$ is the fine-structure constant, and $Z = 29$ is the atomic number of the copper. Note that we take copper as an example here since it can satisfy the half-life requirement of the model.
This model can explain the MeV emission line of GRB 221023A well with reasonable parameters.  We assume the half-opening angle $\theta_j$ of the GRB jet is about $0.1$~rad and the redshift $z = 0.1$, we estimate that the jet beaming-corrected gamma-ray emission energy of GRB~221023A is about $7 \times 10^{49}$~erg. 
Considering the propagation distance $d = 4.8 \times 10^{11} ~\rm cm$, which corresponds to the timescale without an observable MeV emission line at the beginning, the Lorentz factor of the jet, $\Gamma = (1 + z) E_{\rm line} / \epsilon_z$, is about $300$. The total mass of heavy nuclei entrained in the GRB jet is $M_{\rm tot, nuclei} = \theta_j^2 c m_i \mathcal{E}_{\rm line} / (4 E_{\rm line} \Gamma_e d \zeta_i)$, approximately $10^{26} {\rm~g}$. Here, $\mathcal{E}_{\rm line}$ is the observed isotropic total energy of the emission line, $\Gamma_e$ is the total excitation rate for an electron of the high-Z ion transitioned from the ground in the lab frame, $m_i$ is the mass of the high-Z ion (i.e., the copper), and $\zeta_i = 0.1$ means the mass fraction capable of producing emission lines relative to the mass of all nuclei entrained in the GRB jet~\cite{Wei2024grb}. We find the kinetic energy of heavy nuclei in the jet is about $3 \times 10^{49}$~erg, which is much larger than the energy budget of the observed emission line, approximately $3 \times 10^{48}$~erg.

\clearpage
\section*{Methods}
\subsection{Fermi data analysis.} 
The GBM consists of 12 sodium iodides (NaI) detectors (8 keV$-$1 MeV) and two bismuth germanate (BGO) detectors (20~keV$-$40 MeV)\cite{Meegan2009}, which has three different data types: continuous time (CTIME), continuous spectroscopy (CSPEC) and time-tagged event (TTE). The CTIME data include eight energy channels and have a finer time resolution of 64 ms. The CSPEC data include 128 energy channels, with a time resolution of 1.024 s. The TTE data consits of individual detector events, each tagged with arrival time, energy (128 channels), and detector number\cite{Meegan2009}. We download the GBM data of GRB~221023A from the public science support center at the official Fermi Website \url{https://heasarc.gsfc.nasa.gov/FTP/fermi/data/gbm/triggers/2022/bn221023862/}.

We extracted spectrum by using the TTE data from the brightest (with the smallest angle between this detector and the source object) two NaI detectors (n0, n1) and one BGO detectors (b0). The light curves were extracted using the GBM Data Tools\cite{GbmDataTools}. The  spectral analysis of the Fermi-GBM data was performed using the Bayesian approach package, namely the Multi-Mission Maximum Likelihood Framework (3ML)\cite{Vianello2015}. We selected the GBM spectrum over $\rm 8-900 \ keV$ and $\rm 0.3-30 \ MeV$ for NaI detectors and BGO detector, respectively.   In order to avoid the iodine K-edge at 33.17 keV \cite{Meegan2009}, we ignore the data for the $30-40\ \rm keV$ energy ranges . The background spectrum from the GBM data was extracted from the CSPEC data with two time intervals before and after the prompt emission phase and modeled with a polynomial function of order $0-4$ (Selected background time intervals: $\text{-}130-\text{-}10 \ \rm s,100-200 \ \rm s$).  We have used the Bayesian fitting method for the spectral fitting, and the sampler is set to the dynesty-nested. And we accounted for intercalibration constant factors among NaI and BGO detectors.

\subsection{Spectral fitting.} 
\begin{figure*}
\centering
\includegraphics[width=0.8\hsize]{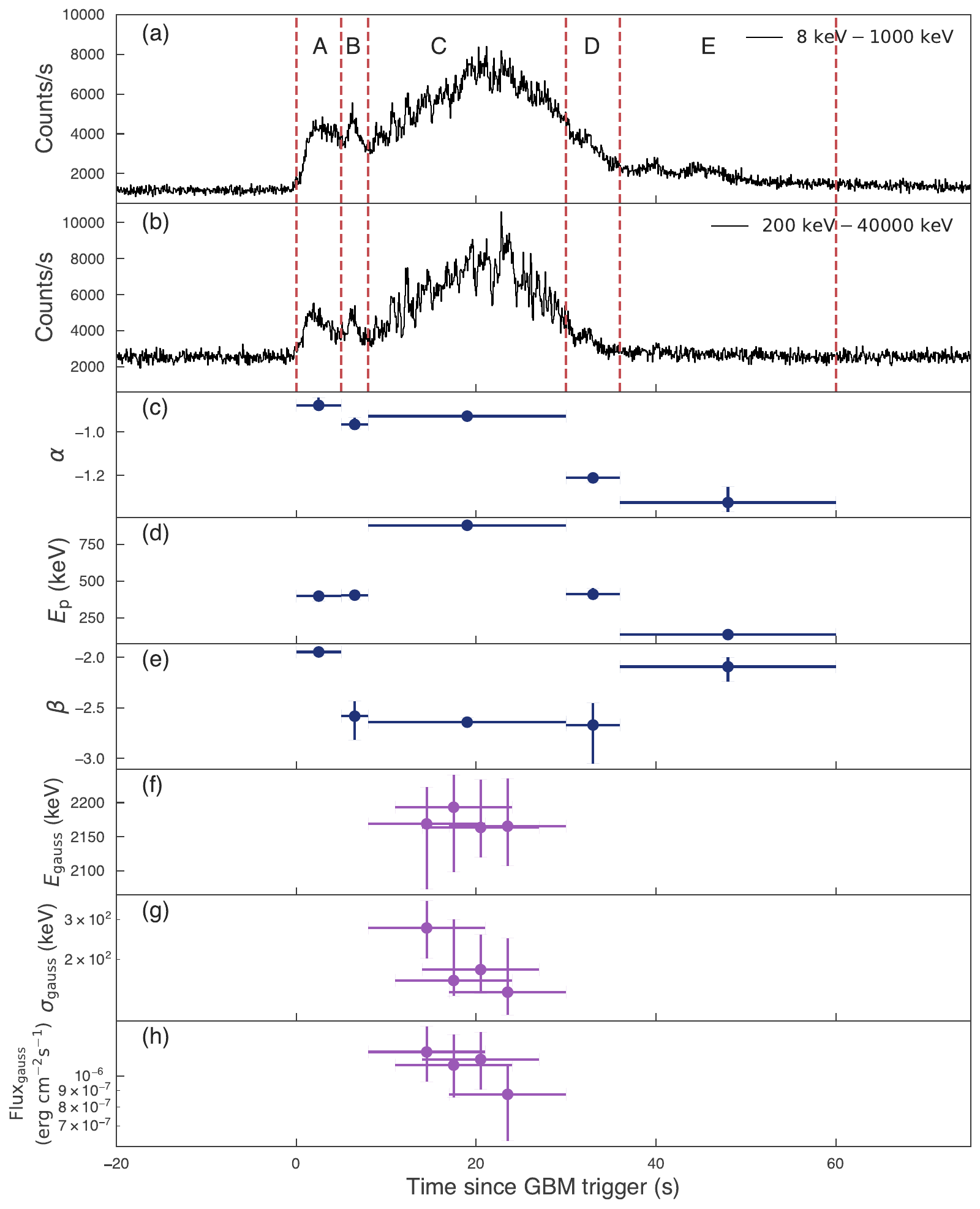}
\caption{\textbf{Multiwavelength light curves and temporal evolution of spectral parameters.} Panels (a) and (b) display the multi-energy band light curves of GRB~221023A observed by Fermi-GBM, with a bin size of 64~ms for each band. The time intervals for spectral analysis are indicated by vertical red dashed lines, labeled as A ($0-5 \ \rm s$), B ($5-8 \ \rm s$), C ($8-30 \ \rm s$), D ($30-36 \ \rm s$) and E~($36-60 \ \rm s$). Panels (c), (d), and (e) show the temporal  evolution of the low-energy spectral index $\alpha$, peak energy $E_p$, and high-energy spectral index $\beta$ of the Band model (deep blue points), respectively. Panels (f), (g) and (h) show the temporal evolution of the central energy $E_{\rm gauss }$, width $\sigma_{\rm gauss }$, and $ \rm Flux_{gauss}$ of the Gaussian component (purple points), respectively. All error bars represent $1\sigma$ uncertainties. Source data are provided as a Source Data file.} 
\label{fig:lightcure}
\end{figure*}

Figure~\ref{fig:lightcure} presents the light curves for GRB~221023A at different energy band. We subdivided the light curve into five intervals labeled A ($0-5 \ \rm s$), B ($5-8 \ \rm s$), C ($8-30 \ \rm s$), D ($30-36 \ \rm s$) and E ($36-60 \ \rm s$),  respectively, which were separated by red dashed vertical lines. We fit the corresponding spectra using the empirical Band function\cite{Band1993}, formulated as follows:
\begin{eqnarray}
N_{\rm Band}(E)=K
\left\{
\begin{array}{lccc}
(\frac{E}{100 \rm~keV})^{\alpha} {\rm exp} (-\frac{E(2+\alpha)}{E_p}), &
({\rm~if}\  E<(\alpha-\beta)\frac{E_p}{2+\alpha}).\\
\left[\frac{(\alpha-\beta)E_p}{(2+\alpha)100 \rm~keV}\right]^{\alpha-\beta}{\rm exp}(\beta-\alpha)(\frac{E}{100 \rm~keV})^{\beta}, &
({\rm~if}\ E\geq (\alpha-\beta)\frac{E_p}{2+\alpha}). \\
\end{array}
\right.
\label{Band model}
\end{eqnarray}
where $K$ is the normalization of Band spectrum, $\alpha$ and $\beta$ are the low and high-energy photon spectral indices, respectively. $E$ is the observational photon energy, and $E_p$ is the peak energy of the $\nu F_{\nu}$ spectrum. The maximum values of the marginalized posterior probability densities and the corresponding $1\sigma$ uncertainties for each parameter of the Band model in each time interval are presented in Table \ref{tab:1}.

The intriguing aspect was the shape of the spectrum in the time interval $8-30 \ \rm s$, as shown in the a and b panels of Figure~\ref{fig:1}, revealing a distinct narrow and bright emission feature between 1~MeV and 3~MeV, this feature did not appear in the other four spectra. We further analyzed the GRBs data detected by the Fermi satellite within ten days before and after the explosion of GRB~221023A. For each of these events, we performed time-resolved spectral analysis using different signal-to-noise ratios and Bayesian blocks, no similar narrow feature were found in these GRBs. In order to  model the narrow emission feature observed at MeV energies, we incorporated a blackbody component into the Band function. However, Blackbody component  is not enough narrow to properly fit the narrow  emission feature. Therefore, we  introduced an additional Gaussian component to fit the spectrum of the time interval $8-30 \ \rm s$. The Gaussian function is defined as follows:
\begin{eqnarray}
N_{\rm gauss}(E) = A \frac{1}{\sigma_{\rm gauss} \sqrt{2 \pi}}{\rm exp}\left({\frac{-(E-E_{\rm gauss})^2}{2~(\sigma_{\rm gauss} )^2}}\right).
\label{gaussian model}
\end{eqnarray}
where $A$ is the normalization of spectrum,  $E_{\rm gauss}$ and $\sigma_{\rm gauss}$ are the central energy and standard deviation of the Gaussian function. We have found that the Gaussian component is well constrained at $E_{\rm gauss}=2154.60_{-65.07}^{+53.37} \ {\rm keV}$, with a width $\sigma_{\rm gauss}=229.36_{-45.29}^{+93.57} \ {\rm keV}$. The fitting results of the spectrum are presented in Table~\ref{tab:1}. The c and d panels of Figure~\ref{fig:1} displays the counts rate and $\nu F_{\nu}$ spectrum, with fitting using the Band function plus a Gaussian component. From the light curve presented in the (a) and (b) panel of Figure~\ref{fig:lightcure}, the MeV narrow emission feature appears during the rising and falling phases of the main pulse. When compared to other time intervals ($0-5 \ \rm s$, $5-8 \ \rm s$, $30-36 \ \rm s$ and $36-60 \ \rm s$), the time interval $8-30 \ \rm s$ exhibits the highest flux and the best signal-to-noise ratio. The evolution of the spectral parameters of the Band function in the best-fit model is shown in the (c), (d), and (e) panels of Figure~\ref{fig:lightcure}. The low-energy spectral index $\alpha$ evolved from -0.88 to -1.33, indicating an evolution from hard to soft.  Additionally, the peak energy $E_p$ varies between $\rm 397 \  keV$ and $\rm 920  \ keV$, showing  the pattern of intensity tracking\cite{LuRuiJing2012}.

For the A ($0-5 \ \rm s$), B ($5-8 \ \rm s$), D ($30-36 \ \rm s$) and E ($36-60 \ \rm s$) time intervals,  we fixed the line width at $\sigma_{\rm gauss} = 200 \ \rm keV$ and the line central energy at $E_{\rm gauss} = 2.1 \ \rm MeV$ in the likelihood fit, thereby deriving the upper limits on the flux of the narrow emission feature, which are $\rm Flux_{\rm gauss} < 5.1 \times 10^{-7} \ \rm{erg \ cm^{-2} \ s^{-1}}$, $\rm Flux_{\rm gauss} < 3.2 \times 10^{-7} \ \rm{erg \ cm^{-2} \ s^{-1}}$, $\rm Flux_{\rm gauss} < 2.4 \times 10^{-7} \ \rm{erg \ cm^{-2} \ s^{-1}}$, and  $\rm Flux_{\rm gauss} < 9.9 \times 10^{-8} \ \rm{erg \ cm^{-2} \ s^{-1}}$, respectively.

\subsection{Model comparison.} 
We employed three different methods to assess the necessity of adding a Gaussian component to the prompt gamma-ray spectrum of GRB~221023A.

The Akaike Information Criterion (AIC) is employed for model comparison when penalizing additional free parameters is necessary to prevent overfitting.  The AIC is formulated as the logarithm of the likelihood with a penalty term \cite{akaike1974,burnham2004}:
\begin{eqnarray}
{\rm AIC} = -2{\rm ln}( \mathcal{L}(d|\theta))  +2\theta .
\label{AiC model}
\end{eqnarray}
where $\mathcal{L}(d|\theta)$ is the likelihood of the model, $\theta$ is the number of free parameters of a particular model. The model with the smallest AIC is favored.  $ \Delta \rm AIC = AIC_{Band}-AIC_{Band+Gaussian}$ provides a numerical assessment of the evidence that model Band+Gaussian is to be preferred over model Band. When $ \Delta \rm AIC>10$, it strongly favors the model Band+Gassian.  As shown in Table~\ref{tab:2}, our results reveal that during the time interval $8-30 \ \rm s$,  the $\Delta \rm AIC$ value reaches its maximum at 51.87, strongly favoring the Gaussian+Band model over the simpler Band model.   In the four finer time-resolved spectra ($8-21 \ \rm s$ (C.1), $11-24 \ \rm s$ (C.2), $14-27 \ \rm s$ (C.3), $17-30 \ \rm s$ (C.4)), the $\Delta \rm AIC$ values vary between 25.76 and 36.55, further strongly favoring the addition of the Gaussian component.

When evaluating the significance of emission or absorption features in spectrum analysis, the Bayesian factor is also a commonly used tool\cite{Freeman1999, Protassov2002, Hurkett2008}. The Bayesian factor is utilized to compare the relative support for different models, serving as a measure to evaluate the strength of evidence in favor of one model over another. Bayesian evidence ($\mathcal{Z}$) is calculated for model selection and can be formulated as follows:
\begin{equation}
		\mathcal{Z} = \int \mathcal{L}(d|\theta) \pi(\theta) d\theta, 
\end{equation}
where $\pi(\theta)$ represents the prior distribution for $\theta$.  The ratio of the Bayesian evidence for two different models is called the Bayes factor (BF). In this paper, the BF is formulated as follows:
\begin{equation}
{\rm BF} =\frac{\mathcal{Z}_{\rm Band+Gaussian}}{\mathcal{Z}_{\rm Band}},
\end{equation}
The corresponding logarithmic expression is as follows:
\begin{equation}
{\rm ln(BF)} = {\rm ln}(\mathcal{Z}_{\rm Band+Gaussian})-{\rm ln}(\mathcal{Z}_{\rm Band}).
\end{equation}
If ${\rm ln(BF)}>8$, it indicates strong evidence in favor of the Band+Gaussian model\cite{jeffreys1998, Thrane2019}. We calculated the Bayes factors for time intervals with narrow emission features (as shown in Table~\ref{tab:2}),  and the results shown that the Band+Gaussian model was preferred in finer time intervals  ($8-21 \ \rm s$, $11-24 \ \rm s$, $14-27 \ \rm s$, $17-30 \ \rm s$) with $\rm ln(BF)$ betweem $2.06-7.34$. Remarkably, during the entire time interval of $8-30 \ \rm s$, the $\rm ln(BF)=9.99$ providing strong statistical support for the addition of the Gaussian component, suggesting the presence of the narrow emission feature.

We also employed the alternative analysis software GTBURST to extract the corresponding spectra from the time intervals exhibiting a narrow emission feature. The extracted spectra were fitted using  the XSPEC 12.11.1 \cite{Arnaud1996},  and the fitting results
similarly indicate the presence of distinct narrow and bright emission feature between 1~MeV and 3~MeV.  $\Delta \chi^2$ represents the statistical difference in the goodness-of-fit between the models Band and Band+Gaussian, the $\Delta \chi^2$ values are displayed in Table~\ref{tab:2}. The highest $\Delta \chi^2$ value of 40.14 was observed in the time interval $\rm 8-30 \ s$, while the $\Delta \chi^2$ values for the other time intervals ranged from 18.53 to 34.49.

\subsection{Background.} 
The selection of time intervals for background subtraction can also impact the analysis of the source spectrum. In order to assess the impact of background subtraction on extracted spectrum. 

In time interval $8-30 \ \rm s$, we calculated the background  spectrum by selecting several different time windows. Even with this approach, the narrow emission features are still clearly visible. We performed both Band and Band + Gaussian fittings in the spectra extracted by subtracting different backgrounds in time interval $8-30 \ \rm s$. As shown in Table~\ref{tab:background}, The central energy $E_{\rm gauss}$ of the narrow emission feature are all around 2.1 MeV and the widths $\sigma_{\rm gauss}$ are all around 200 keV, and the values of the $\Delta \rm AIC$ are around 50. The result of the narrow Gaussian feature is  substantially unaffected.

In four subintervals ($8-21 \ \rm s$ (C.1), ($11-24 \ \rm s$ (C.2), $14-27 \ \rm s$ (C.3), $17-30 \ \rm s$ (C.4)), we extracted the spectra by performing a different selection of the time windows for the background spectrum computation, the background time intervals selected for each time intervals, for $8-21 \ \rm s$: $\text{-}200-\text{-}40\ \rm s,120-250\ \rm s$; for $11-24 \ \rm s$: $\text{-}90-\text{-}10\ \rm s,100-150\ \rm s$; for $14-27 \ \rm s$: $\text{-}90-\text{-}20\ \rm s,180-250\ \rm s$; for $17-30 \ \rm s$: $\text{-}200-\text{-}50\ \rm s,120-250\ \rm s$.

\subsection{Significance calculation of narrow emission feature.} 

\begin{figure*}
\centering
\includegraphics[width=1.0\hsize]{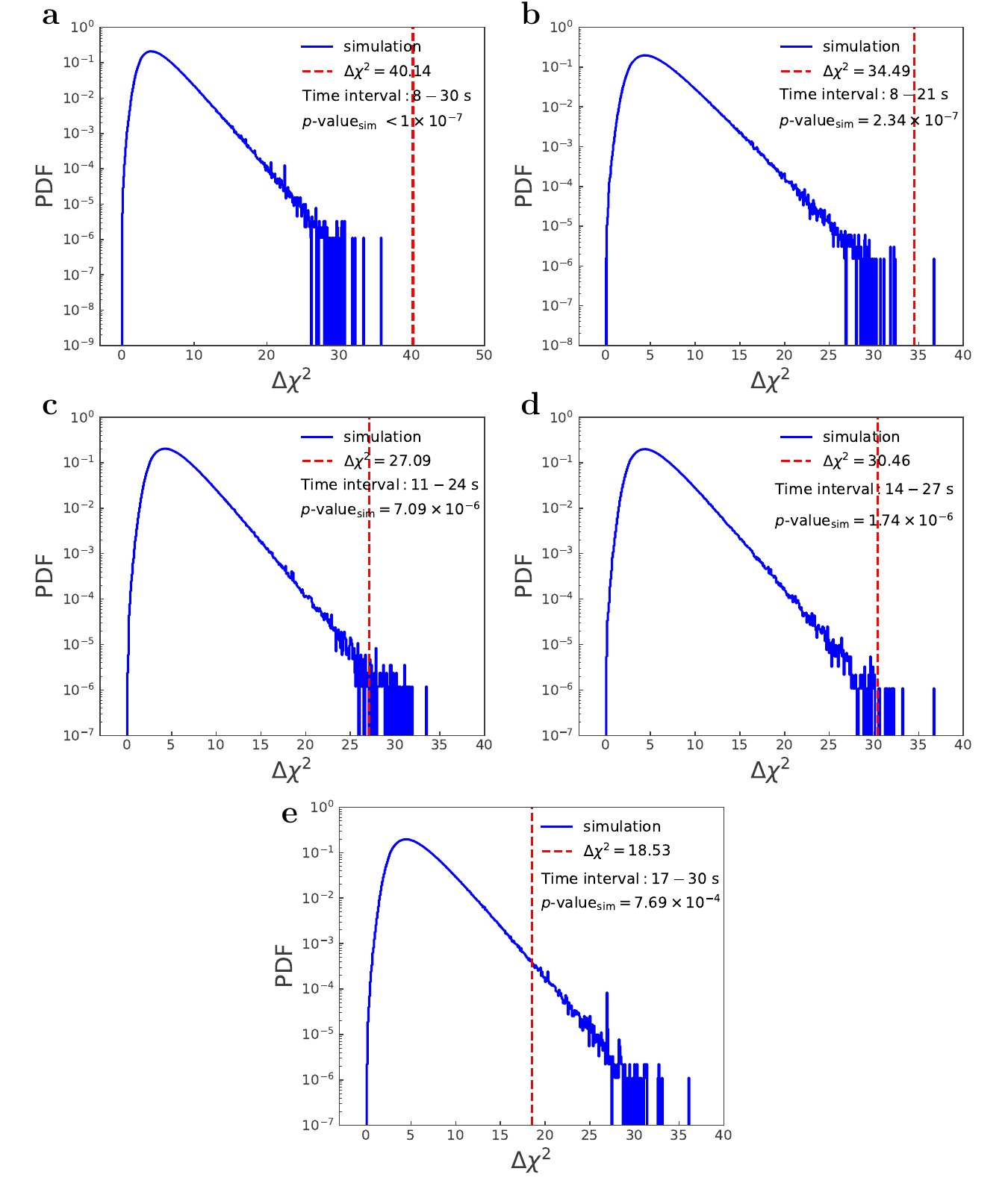}
\caption{\textbf{Simulated $\Delta \chi^2$ distribution.} $\Delta\chi^2$ is the statistical difference in the goodness-of-fit between the models Band and Band+Gaussian. The Panels \textbf{a}, \textbf{b}, \textbf{c}, \textbf{d}, and \textbf{e} show the probability distribution function (PDF) of $\Delta \chi^2$ values obtained from $1 \times 10^7$ simulations for time intervals $8-30 \ \rm s$, $8-21 \ \rm s$, $11-24 \ \rm s$, $14-27 \ \rm s$, $17-30 \ \rm s$, respectively. The red dashed line represents the observed $\Delta \chi^2$ value. The $p$-$\rm value_{\rm sim}$ is the chance probability value obtained from $1.00 \times 10^{7}$ simulations. If after $N$ simulations no $\Delta \chi^2$ value exceeds the actual fitting result, we report $p$-$\rm value_{\rm sim}$ $<1/N$. Source data are provided as a Source Data file.}
\label{fig:t}
\end{figure*}

We calculated the chance probability value ($p$-value) of the narrow emission feature through spectral simulation. The spectral simulation across the entire energy range ($\rm 10 \ keV - 30  \ MeV$) is performed using the fakeit command in XSPEC. These simulations are based on the parameters obtained from fitting the actual data using the Band model. The tclout simpars (based on the covariance matrix at the best fit) command in XSPEC is used to generated randomized model parameters  before each simulation. The total number of spectral simulations $N$ is $1.00 \times 10^{7}$. For each simulated spectra, we perform both Band and Band+Gaussian fittings (search for Gaussian components across the entire energy range of 10~keV to 30~MeV) and record the maximum $\Delta \chi^2$  value~\cite{Protassov2002,Hurkett2008}.
Finally, we assess the significance of the narrow emission feature by analyzing the $\Delta \chi^2$ values recorded in Tables~\ref{tab:2}. The $p$-value represents the fraction of simulated $\Delta \chi^2_{i}$ values that exceeds the observed $\Delta \chi^2$ value: 
\begin{equation}
p \text{-} \mathrm{value}_{\mathrm{sim}}  = n[\Delta \chi^2_{i} \geq \Delta \chi^2]/N.
\end{equation}
If after $N$ simulations we still do not obtain a $\Delta \chi^2$ value exceeding the observed value, we report $p$-$\rm value_{\rm sim}$ $< 1/N$. The probability distribution function (PDF) of $\Delta \chi^2$ values obtained from $1 \times 10^7$ simulations for different time intervals are shown in Figure~\ref{fig:t}.        

In the process of calibrating the $\Delta \chi^2$ test distribution through  simulation, the intensity, location and width of the line, are not fixed to predetermined values but are allowed to vary freely during the fit. This is a standard setup when performing the simulation. The number of independent search trials conducted by dividing multiple time intervals in the time series of different GRBs must be considered (the look-elsewhere effect\cite{Gross2010}). The chance probability value $p$-$\rm value_{\rm sim\text{-}trial}$ after considering the correction for the number of independent search trials on the basis of the $p$-$\rm value_{\rm sim}$ is
\cite{Gross2010,Bringmann:2012vr,Weniger2012}:
\begin{equation}
p \text{-} {\rm value}_{\rm sim\text{-}trial} = 1 - (1 - p \text{-} {\rm value}_{\rm sim})^{t}.
\end{equation}
where $t$ is the number of independent search trials. We searched for GRBs spectral lines from the Fermi-GBM catalog (https://heasarc.gsfc.nasa.gov/W3Browse/fermi/fermigbrst.html) in descending order of fluence\cite{vonKienlin2020ApJ}. We excluded GRB~221009A, which already has identified narrow emission features~\cite{Edvige2023, Zhang2024obs}. The extreme brightness of GRB~130427A and GRB~230307A caused detector pile-up effects, so we excluded the saturated time intervals of $4.5-11.5 \ \rm s$ for GRB~130427A and $3-7 \ \rm s$ for GRB~230307A\cite{Ackermann2014Sci,Dalessi2023GCN}. We searched a total of 9 GRBs, for each burst, time intervals were divided based on BGO light curve signal-to-noise ratio greater than 40. This resulted in a total of 256 searches. Therefore, the number of independent search trials $t=256$.

We found the highest statistical significance of narrow emission feature in the time interval $8-30 \ \rm s$, with the chance probability value $p \text{-} \mathrm{value}_{\mathrm{sim}} < 1 \times 10^{-7}$ obtained from results of $1.00\times10^7$ simulations, corresponding to the Gaussian-equivalent significance $> 5.32\sigma$. Considering the correction for the number of independent search trials, the chance probability value decreases to $p \text{-} {\rm value}_{\rm sim\text{-}trial}<2.56\times10^{-5}$, corresponding to the Gaussian-equivalent significance $> 4.20\sigma$. 
The chance probability values for the other time intervals are shown in Table \ref{tab:2}.

\section*{Data availability}
The Fermi-GBM data for GRB~221023A and GRB~221009A used in this paper are publicly available at \url{https://heasarc.gsfc.nasa.gov/FTP/fermi/data/gbm/triggers/2022/bn221023862/} and \url{https://heasarc.gsfc.nasa.gov/FTP/fermi/data/gbm/triggers/2022/bn221009553/}. These data were obtained from the High Energy Astrophysics Science Archive Research Center (HEASARC) at NASA's Goddard Space Flight Center. The Fermi-GBM Gamma-Ray Burst catalog is available at \url{https://heasarc.gsfc.nasa.gov/W3Browse/fermi/fermigbrst.html}. The datasets generated during and/or analyzed during the current study are available from the corresponding author upon request. Source data are provided with this paper.

\section*{Code availability}
The GTBURST package for analyzing the Fermi-GBM Gamma-Ray Burst data are publicly available at \url{https://fermi.gsfc.nasa.gov/ssc/data/analysis/scitools/gtburst.html}.  XSPEC is available at \url{https://heasarc.gsfc.nasa.gov/xanadu/xspec/}. 3ML is available at \url{https://threeml.readthedocs.io/en/stable/}. Fermi-GBM Data Tools is available at \url{https://fermi.gsfc.nasa.gov/ssc/data/analysis/gbm/gbm_data_tools/gdt-docs/notebooks/Trigdat.html}.

\renewcommand{\refname}{References}
\bibliography{ms}

\section*{Acknowledgements}
We acknowledge the use of the Fermi-GBM data provided by the Fermi Science Support Center.
We thank Zi-Qing Xia, Xiaoyuan Huang, Rui-zhi Yang, Tian-Ci Zheng, Qiao Li, and Chang-Xue Chen for technical support. We also thank Rui-zhi Yang for theoretical insights.
This work was supported by the Strategic Priority Research Program of the Chinese Academy of Sciences (grant No. XDB0550400), the National Key $\rm R \& D$ Program of China (2024YFA1611704), the NSFC (No.~12473049, 12233011, 11921003, 12321003, 12225305).
X.L is supported by the Youth Innovation Promotion Association CAS.
H.N.He is supported by Project for Young Scientists in Basic Research of Chinese Academy of Sciences (No. YSBR-061), and by NSFC under the grants No. 12173091, and No. 12333006.
Y.W is supported by the Jiangsu Funding Program for Excellent Postdoctoral Talent (grant No. 2024ZB110), the Postdoctoral Fellowship Program (grant No. GZC20241916) and the General Fund (grant No. 2024M763531) of the China Postdoctoral Science Foundation.
J. R is support by the General Fund (grant No. 2024M763530) of the China Postdoctoral Science Foundation.
Z.Q.S is supported by the NSFC (No. 12003074).

\section*{Author contributions}
D.M.W and L.Y.J launched the project. L.Y.J, Y.W, X.L and Z.Q.S processed and analyzed the data. H.N.H, D.M.W, L.Y.J, J.R, Y.J.W and Z.P.J contributed to the theoretical interpretations to the event. All authors prepared the paper and joined the discussion.

\section*{Competing interests}
The authors declare no competing interests.

\clearpage
\begin{table}
\centering
 \caption{\textbf{The spectral fitting results for each time interval of GRB~221023A.} $\alpha$ and $\beta$ are the low and high energy spectral indices of the Band function respectively, $E_p$ is the peak energy of the Band function $\nu F_{\nu}$ spectrum. $E_{\rm gauss}$ and $\sigma_{\rm gauss}$ are the central energy and standard deviation of the Gaussian function.
 The energy fluxes are calculated between $10 \ \rm keV$ and $30 \ \rm MeV$, all errors represent the $1\sigma$ uncertainties.}
 \resizebox{\textwidth}{!}{
 \renewcommand{\arraystretch}{1.5}
 \begin{tabular}{cccccccccp{2.3cm}p{2.3cm}}
\hline  
$\mathrm{Time \ interval}$ & $\mathrm{Model}$  & $\alpha$ &  $E_{p}$ & $\beta$  & $E_{\rm gauss}$ &  $\sigma_{\rm gauss}$  & $\mathrm{AIC}$  & $\mathrm{ln}(\mathcal{Z})$  & $\rm Flux_{total}$ & $\rm Flux_{gauss} $  \\
  ($\rm s$) &   &   & ($\rm keV$)  &   &  ($\rm keV$)   & ($\rm keV$)  &    &    & \shortstack{$ \times 10^{-6}$ \hspace{2em} \\ $\rm(erg \ cm^{-2} \ s^{-1})$} & \shortstack{$ \times 10^{-6}$ \hspace{2em} \\ $\rm (erg \ cm^{-2} \ s^{-1})$}\\
\hline  
$\rm 0.00-5.00 \ [A]$ & $\mathrm{Band}$ & $-0.88_{-0.02}^{+0.04}$ &  $397.44_{-33.90}^{+26.19}$  &  $-1.94_{-0.04}^{+0.03}$  & $...$  & $...$ &  $3318.89$  & $-716.12$ & $13.02_{-0.63}^{+0.69}$  & $\hspace{2em}...$ \\
$\rm 5.00-8.00 \ [B]$ &  $\mathrm{Band}$  &  $-0.96^{+0.03}_{-0.02}$ &  $404.08^{+24.31}_{-34.12}$  &  $-2.58_{-0.23}^{+0.15}$   & $...$ & $...$ & $2763.77$  & $-594.95$  & $7.43_{-0.52}^{+0.61}$ & $\hspace{2em}...$ \\
$\rm 8.00-30.00 \ [C]$ & $\mathrm{Band}$  &  $-0.93_{-0.01}^{+0.01}$  &  $917.44_{-16.19}^{+22.06}$  &  $-2.62_{-0.04}^{+0.03}$   & $...$ & $...$ &  $5230.40$  &  $-1129.88$   & $23.35_{-0.32}^{+0.33}$ & $\hspace{2em}...$ \\
{     }    &   \raisebox{-2ex}{\shortstack{$\rm{Band}$ \\ $+$ \\ $\rm{Gaussian}$}} &  $-0.93_{-0.01}^{+0.01}$  & $891.07_{-33.19}^{+3.03}$  &  $-2.65_{-0.02}^{+0.05}$  &  $2154.60_{-65.07}^{+53.37}$  &  $229.36_{-45.29}^{+93.57}$ &  $5178.53$ & $-1119.89$ & $23.50_{-0.33}^{+0.33}$  &  $1.02_{-0.18}^{+0.17}$\\
\ \   $\rm 8.00-21.00 \ [C.1]$  &  $\mathrm{Band}$  &  $-0.90_{-0.01}^{+0.01}$   &   $829.83_{-23.85}^{+20.73}$   &   $-2.56_{-0.05}^{+0.04}$   &  $...$   &   $...$  &  $4459.59$   &   $-970.55$   & $21.09_{-0.41}^{+0.40}$  & $\hspace{2em}...$\\
\ \    {      }    &  \raisebox{-2ex}{\shortstack{$\rm{Band}$ \\ $+$ \\ $\rm{Gaussian}$}}  &  $-0.90_{-0.01}^{+0.01}$  &   $789.83_{-25.03}^{+18.80}$  &   $-2.60_{-0.05}^{+0.05}$  &  $2168.79_{-96.16}^{+53.83}$  & $275.74_{-73.46}^{+88.29}$  &   $4433.83$    &   $-966.68$  &  $21.30_{-0.38}^{+0.30}$  &  $1.21_{-0.24}^{+0.23}$\\
\ \   $\rm 11.00-24.00 \ [C.2]$  &  $\mathrm{Band}$  & $-0.90_{-0.01}^{+0.01}$  & $982.48_{-20.41}^{+26.21}$  &  $-2.58_{-0.05}^{+0.03}$   &  $...$  &  $...$ &  $4697.14$  &  $-1013.51$  & $28.43_{-0.48}^{+0.45}$ & $\hspace{2em}...$ \\
\ \   {    }   &  \raisebox{-2ex}{\shortstack{$\rm{Band}$ \\ $+$ \\ $\rm{Gaussian}$}}  &  $-0.89_{-0.01}^{+0.01}$  &  $949.30_{-37.26}^{+9.67}$  &  $-2.61_{-0.02}^{+0.05}$  &   $2193.24_{-95.04}^{+47.47}$  &  $161.00_{-23.73}^{+138.96}$  &   $4660.67$  &  $-1006.17$  & $28.56_{-0.45}^{+0.48}$ & $1.08_{-0.21}^{+0.25}$ \\
\ \    $\rm 14.00-27.00 \ [C.3]$  &   $\mathrm{Band}$  &  $-0.91_{-0.01}^{+0.01}$  &  $1018.00_{-19.94}^{+27.50}$  &  $-2.58_{-0.04}^{+0.02}$ & $...$  &  $...$  &   $4541.28$   &  $-979.41$  & $31.04_{-0.47}^{+0.47}$ & $\hspace{2em}...$ \\
\ \    {    }  &  \raisebox{-2ex}{\shortstack{$\rm{Band}$ \\ $+$ \\ $\rm{Gaussian}$}}  &  $-0.90_{-0.01}^{+0.01}$  &  $971.13_{-21.38}^{+23.92}$   &   $-2.56_{-0.06}^{+0.01}$  &  $2163.42_{-43.69}^{+70.57}$  &  $180.45_{-38.91}^{+78.01}$ &  $4504.73$  &  $-975.26$  & $31.14_{-0.46}^{+0.49}$  & $1.23_{-0.21}^{+0.23}$\\
\ \  $\rm 17.00-30.00 \ [C.4]$  &  $\mathrm{Band}$  &  $-0.95_{-0.01}^{+0.01}$  &  $1026.30_{-17.21}^{+32.53}$  &  $-2.57_{-0.04}^{+0.03}$  &  $...$  &  $...$  &  $4570.93$  &  $-985.95$   &  $29.25_{-0.48}^{+0.48}$ & $\hspace{2em}...$ \\
\ \    {       }  &   \raisebox{-0.7ex}{\shortstack{$\rm{Band}$ \\ $+$ \\ $\rm{Gaussian}$}}  &   $-0.95_{-0.01}^{+0.01}$  &   $996.46_{-29.22}^{+21.54}$   &   $-2.54_{-0.06}^{+0.07}$  &  $2165.34_{-58.29}^{+69.85}$   &  $143.46_{-29.83}^{+104.96}$  &  $4543.33$  &  $-983.89$  &  $29.40_{-0.48}^{+0.48}$ & $0.88_{-0.25}^{+0.22}$ \\
$\rm 30.00-36.00 \ [D]$  & $\mathrm{Band}$  & $-1.21_{-0.02}^{+0.02}$ &  $409.88_{-29.46}^{+41.10}$  &  $-2.67_{-0.38}^{+0.22}$ &  $...$  &   $...$  & $3398.34$  &  $-733.23$  & $4.66_{-0.33}^{+0.44}$ & $\hspace{2em}...$\\
$36.00-60.00 \ \rm [E]$  &  $\mathrm{Band}$  & $-1.33_{-0.04}^{+0.07}$  &  $135.48_{-21.58}^{+18.45}$  &  $-2.09_{-0.15}^{+0.09}$    &   $...$  &   $...$ & $4644.18$  &  $-1007.68$   &  $1.22^{+0.21}_{-0.19}$ & $\hspace{2em}...$\\
\hline  
\label{tab:1}
\end{tabular}}
\end{table}

\begin{table}
\centering
 \caption{\textbf{The results of evaluating the significance.} $\Delta \rm AIC$ is the AIC value of the Band model minus the AIC value of the Band+Gaussian model. BF is the Bayes factor. $\Delta\chi^2$ is the statistical difference in the goodness-of-fit between the models Band and Band+Gaussian. The $p$-$\rm value_{\rm sim}$ is the chance probability value obtained from $1.00 \times 10^{7}$ simulations. The $p$-$\text{value}_{\text{sim-trial}}$ is the corrected values obtained by accounting for the number of independent search trials based on $p$-$\text{value}_{\text{sim}}$.
The values in parentheses correspond to the Gaussian-equivalent significance.}
 \resizebox{\textwidth}{!}{\begin{tabular}{cccccc}
\hline  
$\mathrm{Time \ interval(s)}$ & $\rm 8.00-30.00 \ [C]$ & $\rm 8.00-21.00 \ [C.1]$ & $\rm 11.00-24.00 \ [C.2]$ & $\rm 14.00-27.00 \ [C.3]$ & $\rm 17.00-30.00 \ [C.4]$ \\
\hline
$\Delta \rm AIC$ & $51.87$ & $25.76$ & $36.47$ & $36.55$  & $27.60 $
\\
$\rm ln(BF)$ & $9.99$ &  $3.87$ & $7.34$ & $4.15$ & $2.06 $
\\
$\Delta\chi^2$ & $40.14$ & $34.49$ &  $27.09$ & $30.46$ & $18.53$
\\
$p$-$\rm value_{\rm sim}$ & $<1.00 \times 10^{-7} $ & $1.00 \times 10^{-7}$ & $6.30\times10^{-6}$ &  $1.00\times10^{-6}$ & $7.69\times10^{-4}$
\\
{    }  & $ (>5.32 \sigma)$ & $(5.32 \sigma)$ & $(4.51 \sigma)$ &  $ (4.89\sigma)$ & $ (3.36\sigma)$
\\
$p\text{-} \rm value_{\rm sim \text{-} trial}$ & $<2.56\times 10^{-5}$  & $2.56\times10^{-5}$ &  $1.61\times10^{-3}$ & $2.56\times10^{-4}$ & $1.79\times10^{-1}$
\\
{  } &  $(>4.20\sigma)$ & $(4.20 \sigma)$ &  $(3.15\sigma)$ & $(3.65\sigma)$ & $(1.34\sigma)$
\\
\hline
\label{tab:2}
\end{tabular}}
\end{table}

\begin{table}
\centering
 \caption{\textbf{Energy spectrum resulted from different time windows selected for the background region in time interval C ($8-30 \ \rm s$).} $E_{\rm gauss}$ and $\sigma_{\rm gauss}$ are the central energy and standard deviation of the Gaussian function. $\Delta \rm AIC$ is the AIC value of the Band model minus the AIC value of the Band+Gaussian model. All errors represent the $1\sigma$ uncertainties.}
\begin{tabular}{lccc}
\hline  
$\mathrm{Background \ selection \ regions}$  &  
$E_{\rm gauss}$ &  $\sigma_{\rm gauss}$  &  $\Delta \rm AIC$  \\    
\ \ \ \  \ ($ \rm s$) & ($\rm keV$) &  ($\rm keV$) 
\\
\hline
$\text{-}98-\text{-}20,90-180$ & $2146.06_{-53.45}^{+67.56}$  &  $189.89_{-9.18}^{+128.11}$ & $48.48$ 
\\
$\text{-}150-\text{-}60,150-200$ &   $2175.97^{+18.86}_{-79.49}$ & $210.85^{+80.14}_{-47.54}$ & $57.71$
\\
$\text{-}290-\text{-}10,85-300$ &  $2158.45_{-61.89}^{+49.76}$  &   $202.85_{-36.32}^{+99.60}$ &  $51.58$
\\
\hline
\label{tab:background}
\end{tabular}
\end{table}

\clearpage

\end{document}